Discussion of the Paper "SPE-187038-MS: Fracture Closure Stress: Reexamining Field and Laboratory Experiments of Fracture Closure Using Modern Interpretation Methodologies"

Mark McClure

ResFrac Corporation

Posted to arXiv Geophysics[1]

April 15, 2019

**Abstract**

In recent years, there has been discussion in the literature regarding methods of estimating the magnitude of the minimum principal stress from subsurface fracture injection tests, commonly called Diagnostic Fracture Injection Tests (DFITs). McClure et al. (2016) used modeling and mathematical analysis to propose that changes should be made to common interpretation techniques. Subsequently, Craig et al. (2017) attempted to refute those findings using field and laboratory observations. This discussion paper reviews the interpretations from Craig et al. (2017) and concludes that their conclusions are unsupported by the data presented.

**Background**

Fracture injection tests are used to estimate the magnitude of the minimum principal stress, Shmin (Hubbert and Willis, 1957; Haimson and Fairhurst, 1967; Hickman and Zoback, 1983). Sometimes, the initial shut-in pressure is assumed to be a good estimate for Shmin. However, this is not always a good assumption. The concept of 'fracture closure pressure' was developed to provide an alternative method for estimating Shmin. As explained by Castillo (1987), under certain idealized conditions, a plot of pressure versus the square root of shut-in time or versus G-time should form a straight line after shut-in and then deviate from a straight line at the point of fracture 'closure.' The closure pressure is typically assumed to be equal to Shmin.

In fracture injection tests in horizontal wells in low permeability rock (nanodarcy to microdarcy), the pressure transient deviates significantly from this theoretical expectation. There is not usually a straight line on a plot of pressure versus G-time. Typically, dP/dG starts high, decreases to a minimum, rises to a maximum, and then gradually approaches zero as shut-in time goes to infinity (Figure 1). This behavior presents challenges for conventional interpretation. The technique of Barree et al. (2009) is widely used to interpret these tests. A plot is constructed by pressure versus G*dP/dG and 'closure' is picked at the point where a linear plot from the origin is tangent to G*dP/dG. This is the so-called 'tangent method.' It is assumed that this closure pick is equal to Shmin.

---

[1] This discussion is adapted from a blog post from October 2017.

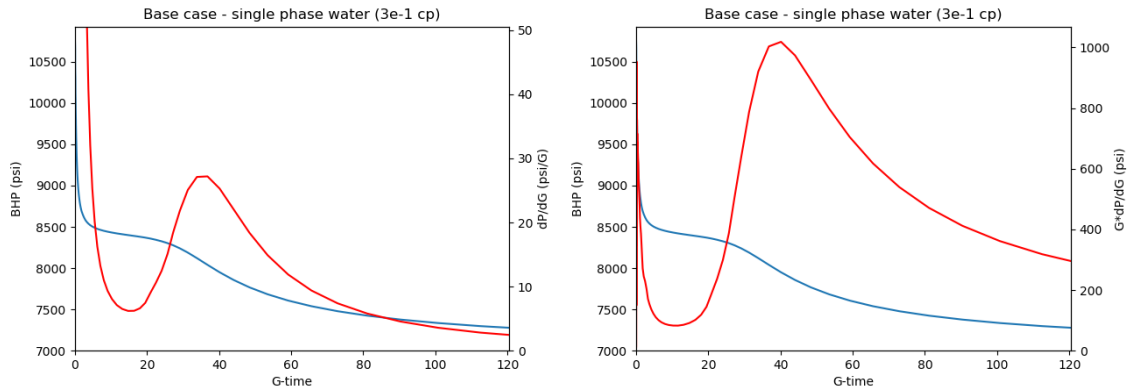

*Figure 1: Example of dP/dG and G\*dP/dG plots from a generic DFIT simulation.*

In low permeability rock, the dimensionless fracture conductivity (conductivity divided by half-length times permeability) remains high after the walls contact. This leads to qualitatively different behavior than in tests in higher permeability formations. All the discussion in this paper considers solely tests performed in low permeability rock.

McClure et al. (2016) performed a modeling study using a detailed physics simulator with fracture propagation, the contacting of fracture walls, and leakoff calculations that take into account changes in fracture pressure over time. When the tangent method was applied to the synthetic data, it did not accurately estimate Shmin. Inspired by these findings, McClure et al. (2016) proposed an alternative method of estimating stress, the 'compliance method.' The compliance method can be explained by writing a chain rule decomposition of the derivative dP/dG:

$$\frac{dP}{dG} = \frac{dP}{dV}\frac{dV}{dG} \sim \frac{leakoff\ rate\ with\ respect\ to\ G-time}{system\ compliance} \qquad (1)$$

Equation 1 shows that dP/dG will increase if the system compliance decreases. When the fracture walls come back into contact, the fracture stiffens due to the solid-to-solid contact of the fracture asperities. Because of roughness, the aperture (volume of fluid stored in the fracture per area) is not zero when the walls touch. To prevent the walls from interpenetrating, the fracture stiffness must asymptotically approach infinity. The implication is that when the walls contact, the compliance drops and the derivative increases. The overall system compliance does not ever go to zero because of wellbore storage.

The derivative dP/dG subsequently peaks and asymptotically approaches zero because of deviation from Carter leakoff. The G-function is derived under the assumption of Carter leakoff. As pressure in the fracture drops, there is deviation from Carter leakoff, leading to a tendency for dP/dG to decrease.

Based on this analysis, McClure et al. (2016) interpret that the increase in dP/dG corresponds to the contacting of the fracture walls. They point out that due to fracture roughness, pressure may be slightly higher than Shmin when the walls come into contact and recommend subtracting a small number, around 75-150 psi, to estimate Shmin.

McClure et al. (2016) noted that the tangent method lacks theoretical justification and has never been validated by field or laboratory data.

The compliance interpretation systematically arrives at higher stress estimates than the tangent method. Sometimes, the difference between the methods is minimal, but sometimes, the difference is 500 psi or more. The tangent method underestimates stress more significantly in formations where fluid pressure is much lower than Shmin (McClure, 2017).

Craig et al. (2017) perform an analysis of field and laboratory data in an attempt to refute the findings from McClure et al. (2016) and validate the 'tangent method' technique. This discussion identifies problems with their analysis.

**Review and discussion of Craig et al. (2017) tiltmeter interpretations**

Figure 6 from Craig et al. (2017) shows tiltmeter and pressure measurements versus time during fracture reopening during Injection 4b from the MDX field demonstration project. It is problematic to plot pressure and tilt versus time. It is more typical to present this data as tilt versus pressure (such as in Figure 9A-4 from Gulrajani and Nolte, 2000). Because the data is plotted versus time, deflections in the curves (ie, changes in the derivatives of tilt and pressure versus time) may be caused by changes in injection rate, rather than changes in system behavior.

Reopening induces a dramatic drop in system stiffness (or equivalently, an increase in system storage coefficient). On a plot of tilt versus pressure, reopening should manifest as a change in the derivative of tilt with respect to pressure (as shown in Figure 9A-4 from Gulrajani and Nolte, 2000).

Because of how Craig et al. (2017) plot the data, interpretation is hampered by uncertainty regarding injection rate. However, even if we assume that injection rate is constant over time, the data does not support their conclusion. Because the paper is copyrighted, the figure is not reproduced below. Instead, the results have been reproduced from their Figure 6c by tracing over the curve with a stylus. The tilt is measured as 'negative' so the curve goes down as the crack opens. The figure from Craig et al. (2017) also shows the concurrent fluid pressure, but in Figure 2, only tilt is shown, for clarity. Figures 6a and 6b from Craig et al. (2017) are qualitatively the similar, and so are not reproduced.

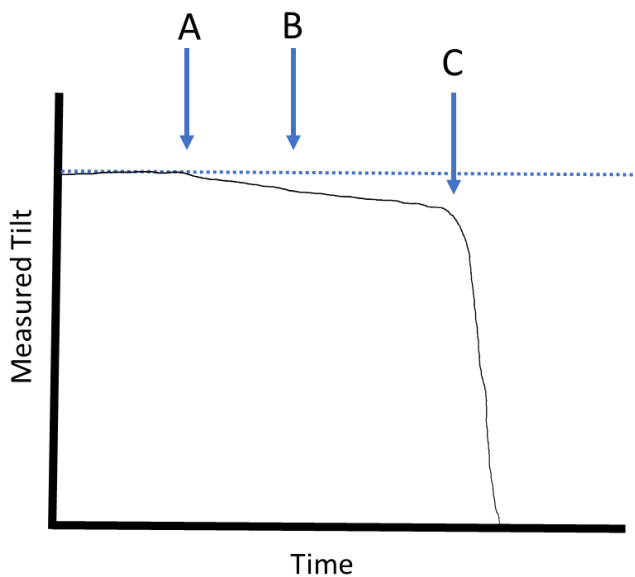

*Figure 2: Reproduction of Figure 6c from Craig et al. (2017). Tiltmeter measurements during reopening in Injection 4b from the MDX field demonstration project.*

There are two points with apparent deflection in tilt: Points A and C. Point A corresponds to 2500 psi, lower than the predicted reopening pressure from either the tangent or compliance methods. Point C corresponds to the point when pressure is equal to the closure pressure from Injection 4a, as interpreted by the compliance method (3200 psi). The fluid pressure at Point B corresponds to the reopening pressure predicted by the tangent method (2900 psi). After Point C, the tilt goes off the scale of the graph as the fracture opens further.

Craig et al. (2017) claim that the data in Figure 2 confirms the tangent method pick that stress is equal to 2900 psi (Point B). However, the curve is straight at Point B. The curve deflects at Points A and C. There is not a discernable deflection at Point B.

Fractures have roughness, and joints have a nonnegligible compliance and ability to store fluid, even when their walls are still in contact (Barton et al., 1985; Jaeger et al., 2007). Thus, some fluid will seep into the fracture prior to the full separation of the walls, followed by a much larger change in aperture once the walls fully separate. Point A corresponds to the point where nonnegligible amount of fluid seeps into the fracture (while the walls are still in contact). Point C may be the point corresponding to full separation of the fracture walls, but again, a true interpretation is impossible because of how the data is plotted.

Figure 15 from McClure et al. (2016) shows tiltmeter data from Injection 2B. This figure shows a plot of tilt versus pressure, making it possible to discern changes in system storage coefficient. The figure shows the fracture reopening at a pressure consistent with the compliance method closure pick from the previous injection and inconsistent with the holistic pick. The data in Figure 15 from McClure et al. (2016) was reproduced from Figure 9A-4 from Gulrajani and Nolte (2000). Gulrajani and Nolte (2000) label the closure/reopening pressure at a pressure is consistent with the compliance method pick from the previous injection and inconsistent with the holistic pick. Gulrajani and Nolte (2000) note that a step-rate test was performed in this well, and that result was consistent with the higher value of closure pressure (which is consistent with the compliance method interpretation and inconsistent with the holistic analysis). Craig et al. (2017) do not mention these other published references who looked at the same data and came to the opposite conclusion, except to speculate that the step-rate test may be unreliable because the pressures response was not 'smooth' enough.

In their Figure 14, Craig et al. (2016) provide data from a different test, the "C" sands. The y-axis scale on their plots do not show the fluid pressure is at the point when the tilt begins to sharply increase as the crack reopens. However, the plots do clearly show that when the fluid pressure reaches their 'holistic' closure pick, there is no apparent deflection in tilt, and that there is a very large, abrupt reopening of the fracture later, at a much higher pressure. Their Figure 14b actually shows the fracture experiencing gradual *closure* at their predicted reopening pressure (probably due to effects from the previous injection cycles), and then later, a sharp and abrupt reopening at a higher pressure. This is evidently a consequence of dynamic poroelastic changes from previous changes, but regardless, it certainly does not validate the tangent interpretation. Again, because Craig et al. (2016) do not provide plots of tilt versus pressure, it is impossible to perform a proper interpretation. But again, even if we assume a constant injection rate over time, the plots contradict their interpretation because they fail to show a deflection in the curve at the pressure predicted by the holistic method.

**Review and discussion of Craig et al. (2017) analysis of laboratory experiments**

Craig et al. (2017) review several laboratory experiments in which hydraulic fractures were created in blocks of rock and then pressure was monitored after shut-in.

Figure 16 from Craig et al. (2017) shows laboratory data in which pressure drops from about 30 MPa at shut-in to the minimum principal stress at 15 MPa. They present other laboratory data that is qualitatively similar. Craig et al. (2017) attempt to interpret this data by constructing a plot of pressure, dP/dG, and G*dP/dG versus G-time. Then, they apply the tangent method to perform an interpretation.

Unfortunately, because of the unusual circumstances of the laboratory test, it is not meaningful to attempt a G-function analysis in this context. The G-function is derived such that pressure versus G-time should make a straight line if: (a) fracture stiffness in constant, and (b) leakoff follows Carter leakoff. But in a non-filtercake forming fluid, Carter leakoff is only valid if fluid pressure in the fracture is constant. This is usually approximately valid in field data (prior to closure) because the net pressure is usually small relative to the difference between the ISIP and the formation fluid pressure. But in Figure 16 from Craig et al. (2017), the fluid pressure drops by half from the ISIP to Shmin. When the fluid pressure decreases so much, there is complete deviation from Carter leakoff far before closure.

When deviation from Carter leakoff occurs, this causes dP/dG to decrease because leakoff is occurring more slowly than anticipated by the G-function derivation. Consistent with this expectation, all of the laboratory data reviewed by Craig et al. (2017) shows dP/dG decreasing strongly prior to reaching Shmin (ie, curving downward G*dP/dG). Craig et al. (2017) call this "pressure dependent leakoff" (which is a misnomer because all leakoff is pressure dependent), using this term to describe a period of anomalously high permeability shortly after shut-in that may be caused by dilation of secondary fractures and fissures. In other words, Craig et al. (2017) incorrectly interpret this pressure transient – caused by the decreasing fracture pressure over time – as representing pressure dependent permeability.

This deviation from Carter leakoff has a very strong overprint on the behavior of the transient when plotted on a G-function plot. The deviation from Carter leakoff (decreasing dP/dG) occurs simultaneously with closure (increasing dP/dG). The two effects occur simultaneously in the lab experiments reviewed by Craig et al. (2017), causing a complex behavior from which closure cannot be inferred from a G-function plot. Wang and Sharma (2017) propose a method for inferring fracture stiffness in the case of pressure deviating from Carter leakoff (see their Figures 33 and 37). Because of the strong deviation from Carter leakoff, the method of Wang and Sharma (2017) would be an appropriate method of interpreting the laboratory data, rather than a G-function plot.

Craig et al. (2017) claim this data refutes the compliance method because dP/dG does not increase (equivalent to a curving upward G*dP/dG) at closure. This claim is based on a misapplication of the G-function plotting technique.

Figure 16 from Craig et al. (2017) shows G*dP/dG is flat after the early shut-in period. Craig et al. (2017) claim the tangent method closure pick is located in the middle of this flat line, near the location of Shmin. Picking closure at this point does not conform with the typical application of the tangent method. However, it allows the interpretation to correctly identify the value of Shmin, which was known when the interpretation was performed.

Craig et al. (2017) show one 'blind' test in which they picked closure without knowing Shmin and then afterwards asked the experimentalist to tell them the true value. However, they performed several blind tests and only reported the results from one of them (Hans de Pater, personal communication), presumably the test in which their estimate was closest to the correct value.

**Conclusions**

To summarize, the conclusions of Craig et al. (2017) are based on flawed and incomplete analysis. The paper: (1) plots tilt versus time, rather than versus pressure, making interpretation impossible, (2) picks 'reopening' in the middle of straight lines on the tiltmeter plots, (3) does not mention previously published analyses that arrived at different conclusions, (4) does not recognize that fractures can accept fluid and experience slight unloading even while their walls are in contact, (5) attempts a G-function analysis of laboratory data with complete deviation from Carter leakoff, which makes G-function interpretation impossible, (6) misinterprets deviation from Carter leakoff due to pressure decrease as being caused by pressure dependent permeability, (7) picks tangent method closure in the middle of a straight, flat line on the G*dP/dG plot (enabling a match to a known 'correct' answer), and (8) presents results from a 'blind' test but does not divulge that other blind tests were performed in which the results were not shown.

Beyond these technical and methodological problems, the interpretations of Craig et al. (2017) appear to be based on a fundamental misunderstanding of the physics. They assume that cracks are either 'mechanically open' or have negligible aperture. This assumption is not consistent with the geometry of real fractures. Recent in-situ core-across studies underscore this, showing remarkable complexity and nonplanarity even at the core scale (Gale et al., 2018). Laboratory studies show the effect of nonlinear joint stiffness after the walls have come into contact (Barton et al., 1985).

The alternative theory, the compliance method interpretation, is based on a mathematical analysis that considers the residual roughness and aperture of the fractures after the walls come back into contact. McClure et al. (2016) show that it is impossible to provide a self-consistent mathematical description of fracture closure that resembles real data without considering this effect. The compliance method (McClure et al., 2016) accounts for fracture roughness, is backed by a succinct mathematical justification, and is validated by field data.

**References**

Barree, R. D., V. L. Barree, and D. P. Craig. 2009. Holistic fracture diagnostics: Consistent interpretation of prefrac injection tests using multiple analysis methods. SPE 107877. Paper presented at the SPE Rocky Mountain Oil & Gas Technology Symposium, Denver, CO.

Barton, N., S. Bandis, K. Bakhtar. 1985. Strength, deformation and conductivity coupling of rock joints. *International Journal of Rock Mechanics and Mining Sciences & Geomechanics Abstracts* **22** (3): 121-140, doi: 10.1016/0148-9062(85)93227-9.

Castillo, J. L. 1987. Modified fracture pressure decline analysis including pressure-dependent leakoff. SPE 16417. Paper presented at the SPE/DOE Low Permeability Reservoir Symposium, Denver, CO.

Craig, D. P., R. D. Barree, N. R. Warpinski, and T. A. Blasingame. 2017. Fracture closure stress: Reexamining field and laboratory experiments of fracture closure using modern interpretation methodologies. SPE 187038. Paper presented at the SPE Annual Technical Conference and Exhibition, San Antonio, TX.

Gale, Julia F. W., Sara J. Elliott, and Stephen E. Laubach. 2018. Hydraulic fractures in core from stimulated reservoirs: core fracture description of HFTS slant core, Midland Basin, West Texas. Paper presented at the Unconventional Resources Technology Conference, Houston, TX.

Gulrajani, Sunil N. and K. G. Nolte. 2000. Chapter 9: Fracture Evaluation using Pressure Diagnostics. *In* Reservoir Stimulation, eds, Michael J. Economides and Kenneth G. Nolte, Wiley.

Haimson, Bezalel, Charles Fairhurst. 1967. Initiation and extension of hydraulic fractures in rocks. Society of Petroleum Engineers Journal **7** (3), doi: 10.2118/1710-PA.

Hickman, Stephen H., Mark D. Zoback. 1983. The interpretation of hydraulic fracturing pressure-time data for in-situ stress determination. In Hydraulic Fracturing Measurements, ed. M. D. Zoback and B. C. Haimson, 44-54. Washington D.C., National Academy Press.

Hubbert, M. K., D. G. Willis. 1957. Mechanics of hydraulic fracturing. Journal of Petroleum Technology **9** (6): 153-168.

Jaeger, J. C., Neville G. W. Cook, Robert Wayne Zimmerman. 2007. Fundamentals of Rock Mechanics, 4th edition. Malden, MA, Blackwell Pub.

McClure, Mark W., Hojung Jung, Dave D. Cramer, and Mukul M. Sharma. 2016. The fracture compliance method for picking closure pressure from diagnostic fracture injection tests. SPE Journal 21 (4): 1321-1339, doi: 10.2118/179725-PA.

McClure, Mark. 2017. The spurious deflection on log-log superposition-time derivative plots of diagnostic fracture-injection tests. SPE Reservoir Evaluation & Engineering 20 (4), doi: 10.2118/186098-PA.

Wang, HanYi and Mukul M. Sharma. 2017. New variable compliance method for estimating in-situ stress and leak-off from DFIT data. SPE 187348. Paper presented at the SPE Annual Technical Conference and Exhibition, San Antonio, TX.